# ON INSTABILITIES OF CONVENTIONAL MULTI-COIL MRI RECONSTRUCTION TO SMALL ADVERSARIAL PERTURBATIONS

Chi Zhang[1,2,*], Jinghan Jia[3,*], Burhaneddin Yaman[1,2,*], Steen Moeller[2], Sijia Liu[4], Mingyi Hong[1], and Mehmet Akçakaya[1,2]

[1]Electrical and Computer Engineering, and [2]Center for Magnetic Resonance Research, University of Minnesota, Minneapolis, MN, United States
[3]University of Florida, Gainesville, FL, United States
[4]MIT-IBM Watson AI Lab, IBM Research, Cambridge, MA, United States

**SYNOPSIS:** Although deep learning (DL) has received much attention in accelerated MRI, recent studies suggest small perturbations may lead to instabilities in DL-based reconstructions, leading to concern for their clinical application. However, these works focus on single-coil acquisitions, which is not practical. We investigate instabilities caused by small adversarial attacks for multi-coil acquisitions. Our results suggest that, parallel imaging and multi-coil CS exhibit considerable instabilities against small adversarial perturbations.

**KEYFINDINGS:** Conventional multi-coil reconstructions are also susceptible to large instabilities from small adversarial perturbations. It is worthwhile to interpret the instabilities of DL methods within this broader context for the practical multi-coil setting.

**INTRODUCTION:** Deep learning (DL) reconstruction has recently received much attention due to its improved reconstruction quality[1-5]. While DL has been transformative in many image processing tasks, it is well-understood that these methods may be susceptible to instabilities arising from small adversarial perturbations due to their non-linear nature[6-8]. Such instabilities were also explored for MRI reconstruction recently[9], which suggested that both researchers and FDA need to be cognizant of these issues. Several follow-up studies[10,11] explored adversarial training frameworks to improve the robustness of DL-MRI reconstruction. However, all these works concentrated on a single-coil setup, which has little practical application.

In this work, we investigate how small adversarial perturbations affect multi-coil MRI reconstruction, particularly using conventional non-DL methods. Our results indicate that for multi-coil MRI reconstruction, parallel imaging and multi-coil compressed sensing (CS) methods are also susceptible to large instabilities from small adversarial perturbations.

**METHODS**:
Multi-coil MRI Acquisition Model and Inverse Problem: The multi-coil encoding model is given as
$$\mathbf{y}_\Omega = \mathbf{E}_\Omega \mathbf{x} + \mathbf{n}, \tag{1}$$
where $\mathbf{y}_\Omega$ is the acquired measurements with sub-sampling pattern $\Omega$, $\mathbf{E}_\Omega$ is the multi-coil encoding matrix, and $\mathbf{n}$ is noise. For i.i.d. Gaussian noise, the maximum likelihood estimate is
$$\arg\min_{\mathbf{x}} \|\mathbf{y}_\Omega - \mathbf{E}_\Omega \mathbf{x}\|_2^2 = (\mathbf{E}_\Omega^H \mathbf{E}_\Omega)^{-1} \mathbf{E}_\Omega^H \mathbf{y}_\Omega, \tag{2}$$

which is the CG-SENSE[12] output without regularization. Alternatively, a regularized version of the problem can be solved
$$\arg\min_{\mathbf{x}} \|\mathbf{y}_\Omega - \mathbf{E}_\Omega \mathbf{x}\|_2^2 + \mathcal{R}(\mathbf{x}) \tag{3}$$
where $\mathcal{R}(\cdot)$ is a regularizer, e.g. Tikhonov or $l_1$-norm of transform coefficients. In DL methods that rely on algorithm unrolling[5], the regularizer is implicitly learned through neural networks, leading to a non-linear representation.

Adversarial Attacks: Let $\mathbf{z}_\Omega = \mathbf{E}_\Omega^H \mathbf{y}_\Omega$ denote the zero-filled image, and $f(\mathbf{z}_\Omega)$ be a reconstruction algorithm that takes as input the zero-filled image. Note that while the reconstruction algorithm may take $\mathbf{y}_\Omega$ as input, using the zero-filled image allows consistency with the setup in[9]. We use an $l_\infty$-attack, i.e. the attack $\mathbf{r}$ on the input with $\|\mathbf{r}\|_\infty < \varepsilon$ leads to a reconstruction $f(\mathbf{z}_\Omega+\mathbf{r})$ that largely deviates from the original output $f(\mathbf{z}_\Omega)$. The attack is chosen on the zero-filled image instead of the fully-sampled image as done in[9], because the latter is not practical: 1) One does not have access to fully-sampled images to generate a practical attack, 2) In multi-coil MRI, the encoding operator is not known exactly, but estimated. Finally, the attack is not chosen in k-space, since it is difficult to define an $l_\infty$-perturbation in k-space due to the varying signal strength between center and edges.

Imaging Data and Experiments: Coronal proton density knee MRI with 15-channel coils were obtained from the fastMRI database[13]. An acceleration rate, R = 4 with 24 ACS lines was used, as common in DL-MRI reconstruction[14].

Both uniform and random undersampling were investigated. For the former, CG-SENSE and GRAPPA were considered, while for the latter, CG-SENSE and multi-coil CS were explored. For both undersampling patterns, the attacks were performed on the CG-SENSE solution. Specifically, the CG algorithm was unrolled for 10 iterations[3]. The attack was generated on this unrolled CG using fast gradient sign method[15] with an MSE loss and $\epsilon=\|\mathbf{z}_\Omega\|_\infty/255$ (Fig. 1). For random undersampling, this attack was used directly on multi-coil CS. For uniform undersampling, the small perturbation attack on the zero-filled image was converted to k-space for GRAPPA. Among the infinitely many k-space perturbations on $\Omega$ that led to **r**, the minimum $l_2$ solution was picked.

For testing, CG-SENSE used 10 iterations; GRAPPA used 5×4 kernels. Multi-coil CS reconstruction utilized variable splitting, $l_1$-norm of Daubechies4 wavelets as the regularizer, and its parameters were tuned empirically. All coil maps were generated using ESPIRiT[16].

Additionally, an attack was generated on a pre-trained DL method[17,18] based on an unrolled network to investigate whether the linear data-consistency units or the non-linear neural networks used for regularization were affected more substantially.

**RESULTS:**
Fig. 2 depicts CG-SENSE and GRAPPA results for uniform undersampling. While the perturbation causes no visual difference in the fully-sampled image, both methods fail under the attack. Fig. 3 shows CG-SENSE and multi-coil CS reconstructions for random undersampling. The same conclusions apply, with both methods failing under a non-visible small perturbation.

Fig. 4 depicts the results of an unrolled neural network under an attack that targets it end-to-end. There is no major change when the attack is run through a single regularizer unit, but output collapses when the attack is passed through a single data-consistency unit. This suggests the end-to-end attack on an unrolled network targets the linear data-consistency units.

**DISCUSSION AND CONCLUSIONS:**
Our results indicate that for multi-coil MRI reconstruction, parallel imaging and multi-coil CS are also susceptible to large instabilities from small adversarial perturbations. Moreover, for DL reconstruction that utilize $\mathbf{E}_\Omega$ explicitly, adversarial attacks predominantly target the linear data-consistency units. The ill-conditioning of the encoding operator is well-discussed for CG-SENSE in non-Cartesian acquisitions, which has led to an early stopping criterion in practice[19]. While in general, it is hard to compute the condition number, which depends on the coil configuration and R, adversarial attacks enable a method to exploit such ill-conditioning. Since these attacks also breakdown multi-coil MRI reconstruction methods, including parallel imaging and CS, it is worthwhile to interpret the instabilities of DL methods within this broader context.

**ACKNOWLEDGEMENTS:** *The first three authors contributed equally to this work. Grant support: NIH R01HL153146, NIH P41EB027061, NIH U01EB025144; NSF CAREER CCF-1651825

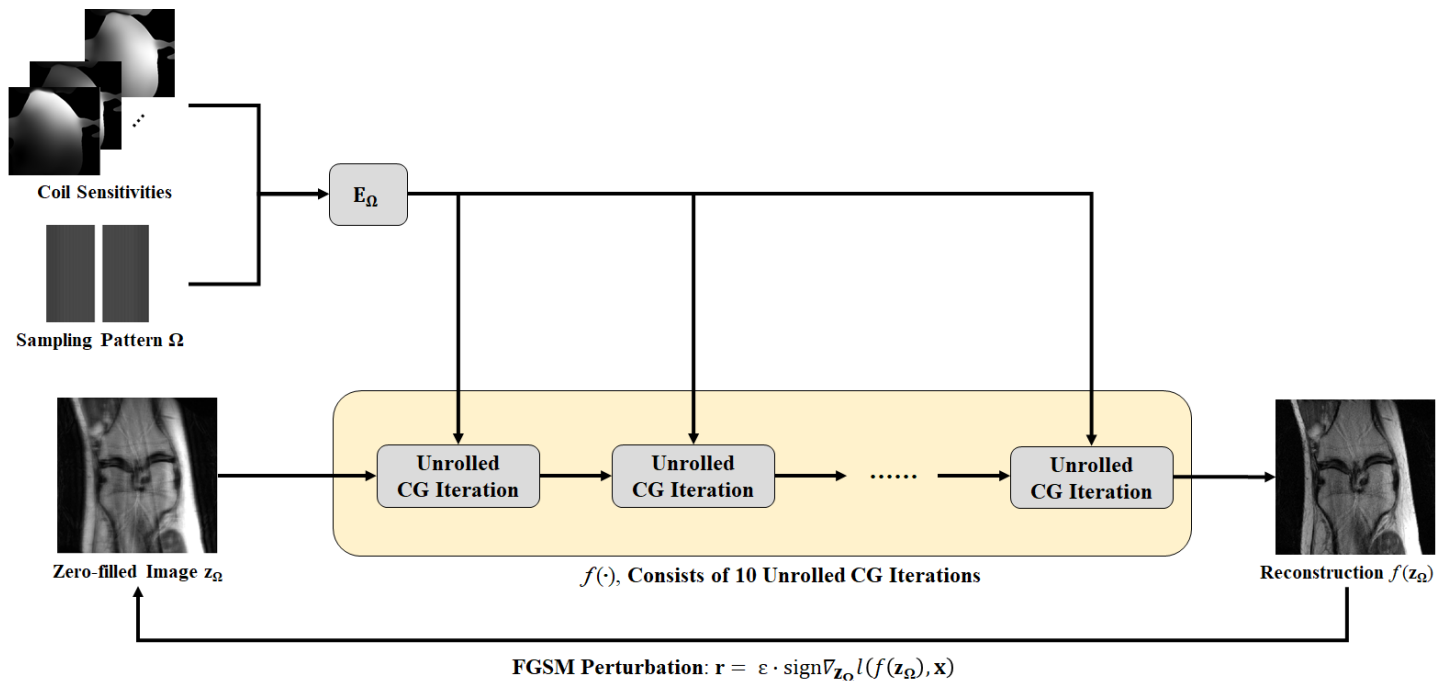

**Figure 1.** The process for generating the small adversarial perturbation using unrolled CG-SENSE. Fast gradient sign method (FGSM) is used to generate the perturbation **r** via a single backpropagation through the unrolled CG algorithm, where $l(\cdot, \cdot)$ denotes MSE loss. For testing, **r** is added to the zero-filled image, $z_\Omega$, which is then run through the relevant reconstruction algorithm.

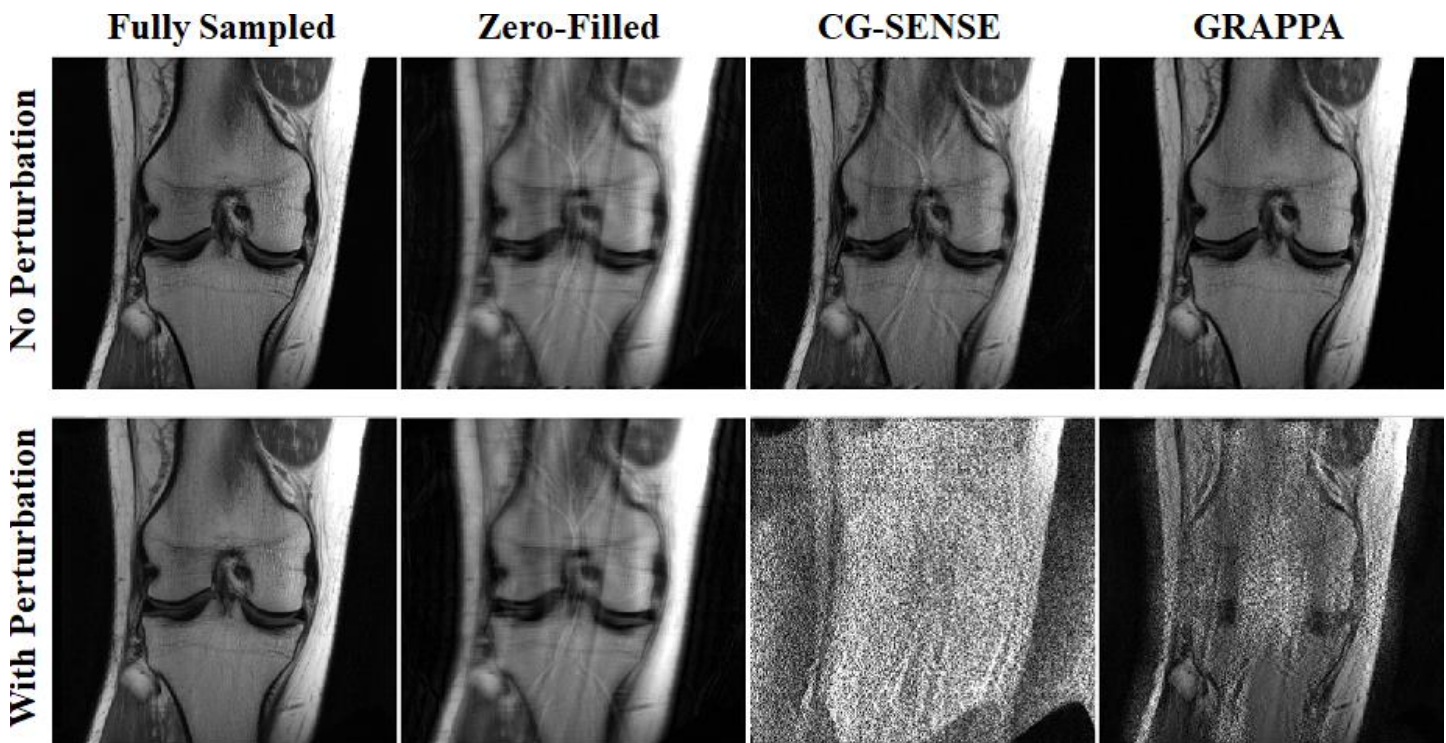

**Figure 2.** Reconstruction results for uniform undersampling without (top row) and with (bottom row) attack. The small perturbation causes no visual difference in the fully-sampled image or the zero-filled image. The reconstructions without attack show some residual aliasing due to the R=4 acceleration. Both CG-SENSE and GRAPPA visibly fail with the small perturbation attack.

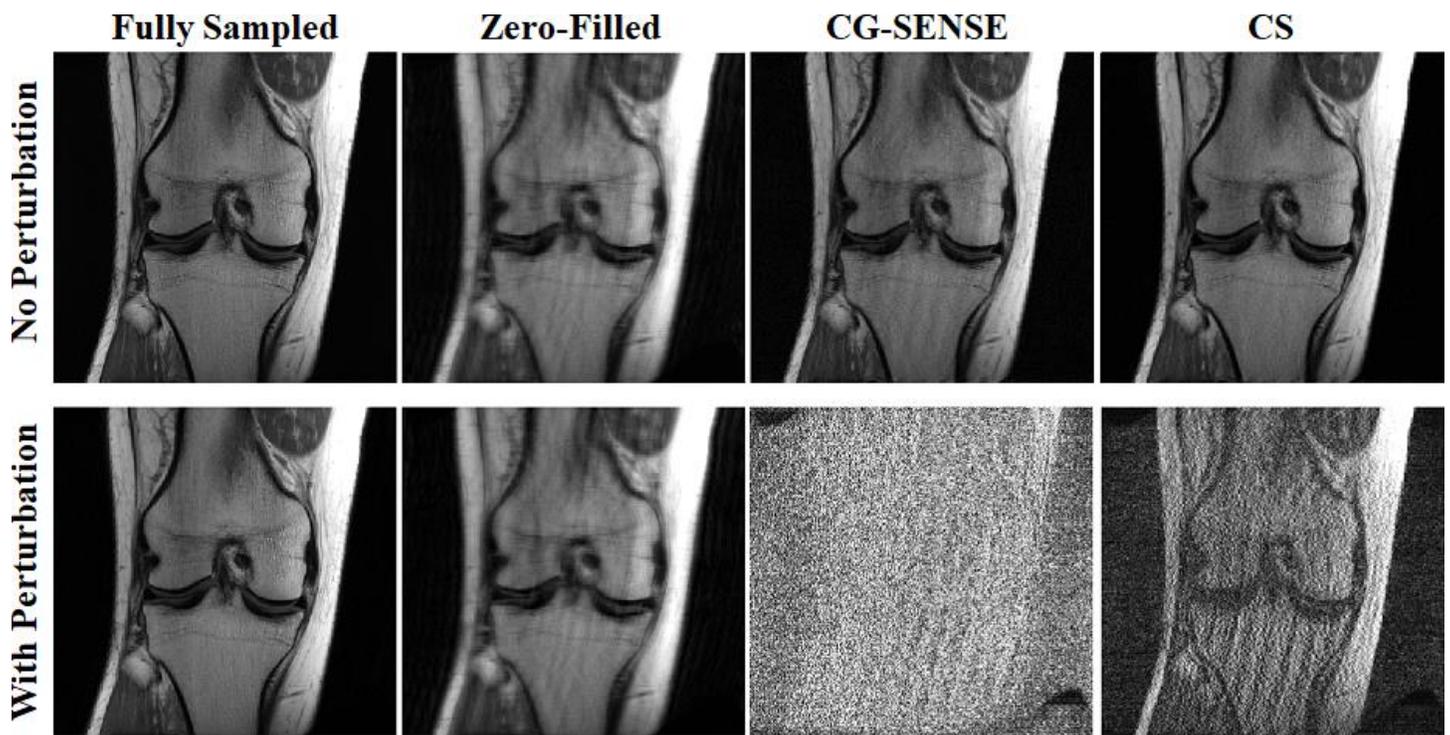

**Figure 3.** Reconstruction results for random undersampling without (top row) and with (bottom row) attack. The small perturbation leads to no visual change in the fully-sampled or zero-filled image. For the input without attack, CG-SENSE has visible and multi-coil CS has subtle aliasing artifacts at R=4. For the attack input, both these reconstructions fail although they are run with the exact same parameters as the non-attack case.

(a)

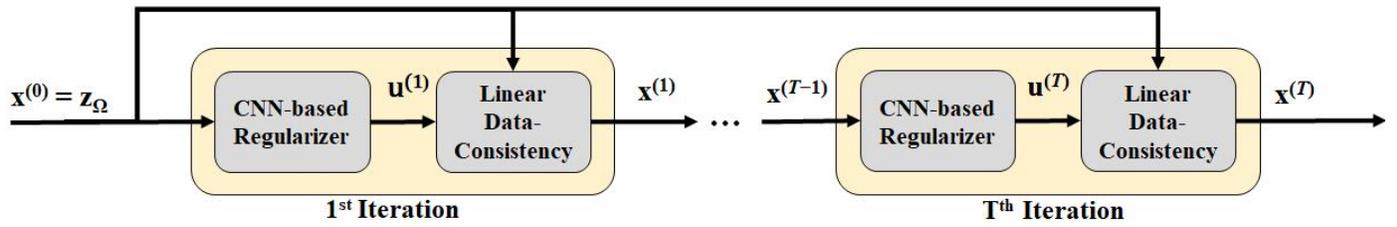

**Regularizer CNN:** $\mathbf{u}^{(i)} = \arg\min_{\mathbf{u}} \mu \|\mathbf{x}^{(i-1)} - \mathbf{u}\|_2^2 + \mathcal{R}(\mathbf{u})$

**Linear Data-Consistency:** $\mathbf{x}^{(i)} = \left(\mathbf{E}_\Omega^H \mathbf{E}_\Omega + \mu \mathbf{I}\right)^{-1} \left(\mathbf{z}_\Omega + \mu \mathbf{u}^{(i)}\right)$

(b)

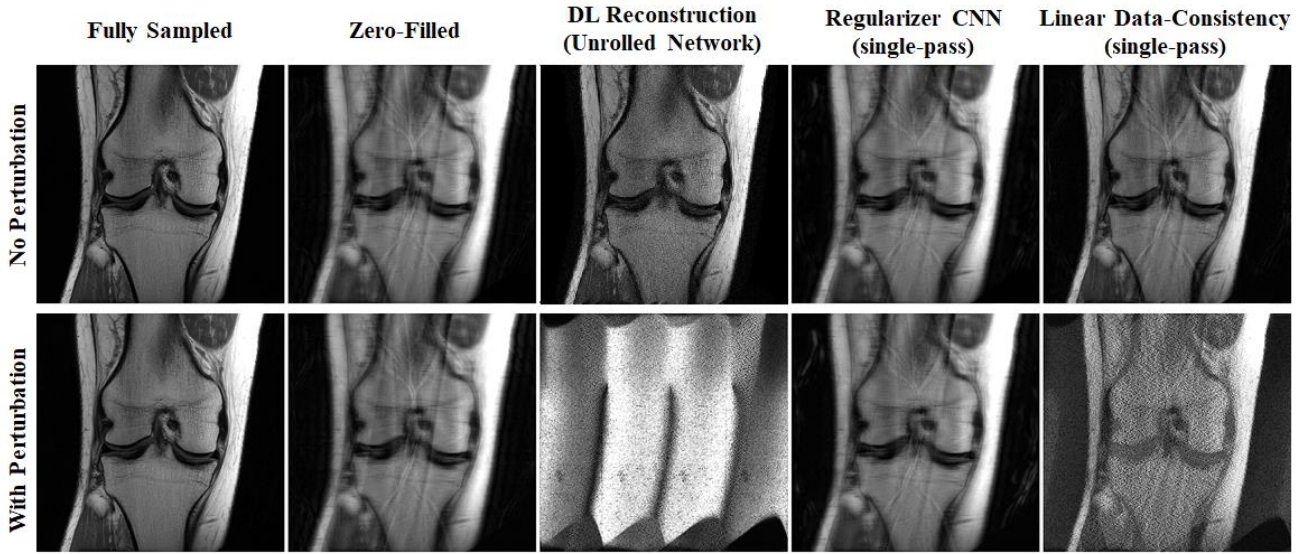

**Figure 4. (a)** Schematic of the unrolled network, with ResNet regularizer and linear data-consistency (DC) unit. **(b)** DL results for uniform undersampling. Perturbation does not visually alter fully-sampled or zero-filled images. Without attack, DL leads to good quality. With the attack, DL collapses, as with conventional methods in Fig. 2&3. The attack through a single CNN regularizer (4th col.) shows no alteration, but it shows visible degradation through a single data-consistency (5th col.), suggesting the attack targets linear DC units more than CNN regularizers in the unrolled network.